\documentclass{article}
\usepackage{axodraw,pstricks,graphicx,epsfig,color,amssymb,amsmath,amscd}

\newcommand{\be}{\begin{eqnarray}}
\newcommand{\ee}{\end{eqnarray}}
\newcommand{\bi}{\bibitem}

\newcommand{\rar}{\rightarrow}
\newcommand{\lrar}{\leftrightarrow}

\newcommand{\rp}{${\cal R}$-parity }

\def\mss{M_{SUSY}}
\def\mpl{m_{Pl}}

\begin{document}

\title{
COSMOLOGICAL CHARGE ASYMMETRY 
AND RARE PROCESSES IN PARTICLE PHYSICS
}
\author{ A.D. Dolgov\\
{\it  Istituto Nazionale di Fisica Nucleare, Sezione di Ferrara,}\\
{\it     I-44100 Ferrara, Italy}\\
{\it Dipartimento di Fisica, Universit\`a degli Studi di Ferrara,}\\
{\it       I-44100 Ferrara, Italy}\\
{\it Institute of Theoretical and Experimental Physics,}\\
{\it       113259, Moscow, Russia}
}
\maketitle

\baselineskip=11.6pt

\begin{abstract}

Two scenarios of low temperature baryogenesis in theories with TeV 
scale gravity are discussed. It is argued that strong gravity at TeV 
energies is very favorable for baryogenesis. In both scenarios the 
proton decay is either absent or suppressed far below existing bounds. 
On the other hand, neutron-antineutron oscillations are at the verge 
of discovery. Some other rare decays with non-conservation of lepton
or baryon numbers are predicted.
\end{abstract}

{It is experimentally established that neither {baryonic} nor
individual {leptonic} numbers are conserved.}
Neutrino oscillations are known to mix 
{electronic, muonic, and tauonic} neutrinos, resulting in
nonconservation of all these quantum numbers. On the other hand,
{astronomy proves that} 
{baryon number is not conserved.} One can say that
{since we exist, baryons are not forever.} Indeed a suitable for
life universe canot be created if baryonic charge were conserved.
The chain of arguments goes as follows. First, the asronomical data
strongly suggest that
{inflation is an ``experimental'' fact.} There are many reasons to 
believe that this is true:\\
1. We do not know any other way to make the observed universe.\\
2. It explains the origin of expansion.\\
3. It solves the problems of homogeneity, isotropy, flatness and
{predicts $ {\Omega =1}$.}\\
4. Inflation creates density perturbations with the {observed} spectrum.

The next important statement is that
{inflation is impossible with conserved baryons.} Inflation could be
realized if the total cosmological energy density is (almost)
constant. However, if baryons are conserved the energy density 
might stay constant at most during 4-5 Hubble times, while for 
successful inflation at least 60 Hubble times are necesary. For
more details see e.g. review~\cite{ad-bs}.

If baryon and lepton quantum 
numbers are not conserved one should 
expect this nonconservation to manifest 
itself in particle physics. The 
well known phenomena searched for are the following:
{unstable proton,} {${( n - \bar n)}$-oscillations,} some
rare decays, as e.g.  ${\mu\rar e\gamma}$, and 
{similar decays of heavier quarks with} 
{$ B$ or $ L$ nonconservation.}
Yet nothing is observed. Though cosmology predicts non-conservation
of baryons and consequently a manifestation of this nonconservation in
particle physics, the magnitude of such effects is expected to be
very small or, at best, unknown 
because the energy scale of cosmological baryogenesis is normally
much higher than that available in terrestrial experiments and, 
what's more, there is usually no direct relation between physics
of baryogenesis and proton decays or neutron-antineutron oscillations.

Here we will discuss some new scenarios of
baryogenesis which explain the observed baryon asymmetry of the
universe and lead to observable consequences in particle physics.
My talk is based on the works made in collaboration with 
F. Urban~\cite{ad-fu} and C. Bambi and K, Freese~\cite{bdf1,bdf2}. 

Let us first consider a rather conservative scenario based on 
{SUSY with broken \rp}. The 
operators which break \rp and experimental bounds on their coupling
constants are enumerated e.g. in review~\cite{R-rev}. 
Cosmological baryogenesis in this model could proceed
through B-nonconserving decays of 
massive SUSY particles induced by B-nonconserving \rp violating 
operators. If masses of supersymmetric particles are not very large,
e.g. ${\mss\sim}$ TeV, deviations
from thermal equilibrium in the standard cosmology are
negligible:
\be{ { 
\frac{H}{\Gamma} \sim \frac{\mss}{\alpha \mpl}\sim 10^{-14},
\label{H-Gamma}
}}
\ee
where $\alpha \sim 10^{-2}$ is the coupling constant. The baryon
asymmetry would be further suppressed at least by factor $\alpha$
because CP-violation manifests itself only in higher orders 
of perturbation theory.

To obtain a reasonable baryon asymmetry the scale of supersymmetry must
be very high:
\be
{\mss\geq 10^{10}}\,\,{\rm GeV}
\label{m-susy}
\ee
However, in this case effects in
{particle physics would be unnoticeable.}

A possible solution which allows both for successful baryogenesis and
for nonnegligible effects in particle physics is offered by
TeV scale gravity.

{There are two known mechanisms for TeV gravity:}\\
{1. Gravity lives in higher dimensional space, while matter lives in
${ D=4}$\cite{add}. }\\
2. Time variation of $ \mpl$ due to the coupling 
$ {\xi R \phi^2}$~\cite{notari}.
It is assumed that initially (in the early universe) $\phi \sim $ TeV,
and later, but prior to nucleosynthesis, it rises up to the Planck value 
$10^{19}$ GeV.

Both these possibilities are practicaly equally good for cosmological
baryogenesis but in the first case care should be taken on the
potential problems with light gravitinos~\cite{gravitinos}.

{Essential \rp-violating operators have the form:}
\be
{\cal L}_{int} = 
-\frac{1}{2}\,\lambda^{ijk}\left(\tilde{u}^*_i\,\bar{d}_j\,d^c_i+
\tilde{d}^*_k\,\bar{u}_i\,d^c_j+ 
{\tilde d}^*_j\,\bar{u}_j\,d^c_k\right) { + h.c.},
\label{L-int}
\ee
where $i,j,k$ are the flavor indices and $u$ and $d$ are respectively
operators of up and down quarks, tidle denotes a superpartner, and
$\lambda^{ijk}$ is a Yukawa type coupling constant. The color
indices are suppressed. These operators do not conserve baryonic
charge, $B$, by one unit and conserve leptonic charge, $L$. 
Correspondingly proton decay is forbidden but transformations of
baryon into antibaryon and, in particular, neutron-antineutron
oscillations are allowed. Such transformations inside a nuclei
would lead to an energetic annihilation and nuclear decay. 
Correspondingly experimental bounds on nuclear stability allow
to put quite strong constraint on some $\lambda^{ijk}$:
\be
{{\lambda_{112}< 10^{-6}}, \,\,\,
{\lambda_{113}< 10^{-3}}}, 
\label{lam-bound}
\ee
while others could be even well above unity.

Non-zero {${\lambda_{112}} $ would lead 
e.g. to $n\bar \Xi$ - transformation.
To make $ n\bar n$ oscillations out of it a ${ \Delta S =2}$
process is necessary. It is strongly suppressed in MSM but
possibly not so much in a supersymmetric extension, in particular 
in minimal supersymmetric model, MSSM.}

The diagramm which induces $\bar n-n$ transformation through 
$\Delta S=2$ processes has the form:
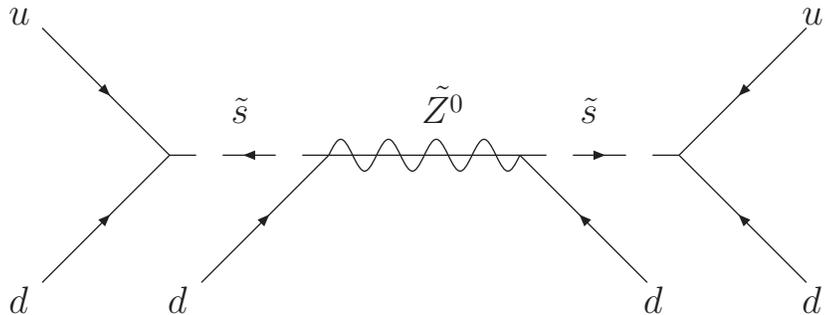
\begin{center}
\fcolorbox{white}{white}{
  \begin{picture}(330,125) (150,-161)
    \SetWidth{0.5}
    \SetColor{Black}
{\bf{ 
   \ArrowLine(162,-53)(210,-101)
    \ArrowLine(162,-149)(210,-101)
    \ArrowLine(222,-149)(270,-101)
    \DashArrowLine(270,-101)(210,-101){10}
    \Photon(270,-101)(342,-101){6}{4}
    \Line(270,-101)(342,-101)
    \Text(234,-89)[lb]{\Large{\Black{$\tilde{s}$}}}
    \Text(210,-161)[lb]{\Large{\Black{$d$}}}
    \Text(150,-161)[lb]{\Large{\Black{$d$}}}
    \Text(366,-89)[lb]{\Large{\Black{$\tilde{{s}}$}}}
    \Text(450,-161)[lb]{\Large{\Black{${d}$}}}
    \Text(390,-161)[lb]{\Large{\Black{${d}$}}}
    \ArrowLine(450,-53)(402,-101)
    \ArrowLine(450,-149)(402,-101)
    \DashArrowLine(342,-101)(402,-101){10}
    \ArrowLine(390,-149)(342,-101)
    \Text(306,-89)[lb]{\Large{\Black{$\tilde{Z^0}$}}}
    \Text(150,-52)[lb]{\Large{\Black{$u$}}}
    \Text(450,-52)[lb]{\Large{\Black{${u}$}}}
}}
  \end{picture}}
\end{center}
Notice that strangeness non-conserving decays of zino, e.g. 
${ \tilde Z \rar \bar u \tilde s }$, are allowed, while  similar
decays of $Z$-boson, not ${ Z \rar \bar u s }$, are not.  
{Both successful baryogenesis and 
${ n \bar n}$ - transformation just above the
 experimental limit might take place.}
For more detail see ref.~\cite{ad-fu}. 

Much more exotic possibility was put forward in refs.~\cite{bdf1,bdf2}
in the frameworks of TeV gravity. The latter is known to suffer from
a serious problem related to nonconservation of all global quantum
numbers. The idea is basically that some (one or a few) particles,
possessing non-zero baryonic, leptonic, or any other global charges,
may form a very dense state inside their common gravitational
radius. In other words, they would form a small virtual black hole.
As is well known, black holes may have ``hairs'' associated only with
conserved quantum numbers related to local (gauge) symmetries, as
e.g. electric charge. On the other hand, if a black hole swallows 
particles with non-zero leptonic, $L$, or baryonic, $B$, charges, it 
immediately ``forgets'' about these charges and may decay into some 
state with zero or any other values of $B$ or $L$. This was first 
observed by Zeldovich~\cite{zeld-bar}, who estimated the life-time
of proton due to formation and decay of a virtual black hole: 
\be{{
\tau_p \sim \mpl^4/m_p^5 \sim 10^{45} \,\,{\rm years}
}}
\label{tau-p}
\ee
This is by far larger than the existing experimental bound
$\tau_p > 10^{33}$ years.
{But if ${ \mpl \sim }$ TeV, ${ \tau_p \sim 10^{-11}}$ s.}
Similar problems exist for ${\mu \rar e \gamma}$ and other rare decays.


These difficulties for low scale gravity was discussed in
ref.~\cite{adams} where it was argued that the fundamental Planck
mass should be much larger than TeV, up to $10^{16}$ GeV. In our
recent works~\cite{bdf1,bdf2} we made an attempt to resolve the
problem of strong gravitational $B$ and $L$ nonconservation 
proposing the so called classical black hole conjecture, which
``dynamically'' forbids an easy formation of black holes (BH).
This conjecture is based on the fact that classical charged and 
rotating black hole can only be formed if it is sufficiently heavy:
\be {
\left( \frac{M_{BH}}{m_{Pl}}\right)^2 > \frac{Q^2}{2} +
\sqrt{ \frac{Q^4}{4} + J^2},}
\label{mass-limit}
\ee
where $Q$ and $J$ are respectively electric charge and angular momentum
of BH. Formally it follows from this expression that if 
$M_{BH}< m_{Pl}$, the black hole can be only electrically 
neutral and non-rotating.
The result (\ref{mass-limit}) is valid for classical black holes and may be 
incorrect for quantum ones. However, physics of quantum black holes is
unknown and one is free to make arbitrary and quite wild 
assumptions. 

In addition to this conjecture of neutral and non-rotating
BHs we impose some, maybe even more questionable, rules 
in calculations/estimates of the amplitude of reactions
with broken global quantum numbers due to virtual BH.
We assume essentially that virtual black holes could be formed
only in s-channel with positive mass (energy) of the created
black hole. Such an assumption and some of the rules which we use
in what follows do not respect many 
usual conditions existing in quantum field theory, in
particular crossing relations between amplitudes. For example,
we allow a virtual BH to decay into, say, a proton and a electron,
but we do not allow a proton to form a scalar BH plus a positron,
with the same amplitude. The picture that we have in mind is a kind 
of time ordering: a BH could be formed in a collision of a neutral system
of particles in the s-channel whereas a BH 
cannot be in the t-channel of a reaction. 
We assume that BHs can be formed out of positive energies of real
particles only and not from virtual energies of particles in closed loops.
For example, BH cannot be formed by vacuum fluctuations, despite the fact
that, according to the standard picture, vacuum fluctuations might 
create a pair or more of
virtual particles both with positive and negative energies.
The mass of the 
BH should be of the order of the energy of incoming (or outgoing)
particles.
In an attempt to describe this in terms of the usual language we come to 
a version of the old non-covariant perturbation theory with all virtual 
particles having positive energies. It corresponds to the choice of only one
mass-shell pole in the Feynman Green's functions. This rule allows
only for BHs with masses which are of the order of the energies of the
initial (or final) particles, as we postulated above.
It may look very strange, to say the least, but virtual BHs are not
well defined objects and we do not know what happens with space-time
at the relevant scales. Taken literally these rules would lead to
violation of some sacred principles of the standard theory (locality, Lorentz 
invariance, and more). Let us remind the reader, however, that the
existing attempts in the literature to invoke virtual BHs are based
on standard quantum field theory in a situation where it is 
almost surely inapplicable.

So it is not excluded that many properties
of the standard field theory are broken, including even Lorentz 
invariance and locality. We cannot of course present any serious
arguments in favor of our construction but it predicts quite
impressive phenomena with clear signatures based on a
very simple set of rules and if these 
effects are discovered, the approach, advocated here, 
may be taken more seriously.
Our goal here is to formulate a reasonable(?) set of rules
which may possibly describe processes with virtual BHs and 
are, at least, not self-contradictory. Based on these rules we 
study phenomenological consequences in particle physics,
which are quite rich and
may be accessible to experiments after a minor increase of accuracy.

{The diagramm that, according to our conjecture, describes 
gravitationally induced proton decay is presented in fig. 1 
Since a 4-body collision is required in order
to form a BH devoid of any quantum number,
the process is strongly suppressed and experimental
constraints can be compatible even with the gravity
scale in TeV range.}
\begin{center}
\begin{figure}
\begin{picture}(340,110)(0,-20)
\Line(95,18)(135,18)
\CArc(135,23)(5,270,360)
\Line(140,23)(140,30)
\CArc(145,30)(5,90,180)
\Line(95,15)(145,15)
\Line(95,12)(135,12)
\CArc(135,7)(5,0,90)
\Line(140,7)(140,0)
\CArc(145,0)(5,180,270)
\ArrowLine(145,35)(185,35)
\ArrowLine(185,35)(235,15)
\ArrowLine(145,15)(185,15)
\Line(185,15)(235,15)
\ArrowLine(145,-5)(185,-5)
\CArc(185,70)(75,270,315)
\PhotonArc(215,30)(30,90,170){3}{4}
\ArrowLine(215,60)(325,75)
\ArrowLine(215,60)(235,15)
\ArrowLine(275,15)(325,35)
\ArrowLine(275,15)(325,-5)
\GCirc(255,15){20}{0.5}
\Text(115,26)[]{$p$}
\Text(157,43)[]{$u$}
\Text(237,45)[]{$e^-$}
\Text(157,23)[]{$d$}
\Text(157,3)[]{$d$}
\Text(192,62)[]{$\gamma$}
\Vertex(235,15){2}
\Vertex(275,15){2}
\Vertex(215,60){2}
\Vertex(185,35){2}
\Text(255,-15)[]{$BH$}
\Text(310,63)[]{$e^+$}
\Text(320,43)[]{$e^-$}
\Text(320,10)[]{$\mu^+$}
\end{picture}
\label{f-proton}\\
Fig. 1. Gravitationally induced proton decay through non-charged and
nonrotating black hole. 
\end{figure}
\end{center}
Similar graphs give rise to $\mu \rar e \gamma$ - decay and other rare
processess with violation of lepton flavor and baryonic numbers. The
results of our calculations, according to ref.~\cite{bdf1} are
collected in Table 1, where the lower bound on the fundamental 
gravity scale is presented for different numbers, $n$, of extra dimensions.

\begin{center}
\begin{tabular}{||c|c|c||}
\hline \hline
 & &  \\
{Process} & {Experiment} & {${ M_*}$, 
${ n=2\, (7)}$} \\
 & &  \\
\hline \hline
& &  \\
{ ${ p \rar eee}$} & {$\quad$ 
$ {\tau > 10^{33}}$ yr $\quad$} 
& {$>$ 2 (8)} \\
\hline
 & &  \\
 ${\mu \rar\gamma e}$ & ${ BR < 10^{-11}}$ & $ >$ 1 (10) \\
\hline
 & &  \\
{ ${\mu \rar eee}$} & {${ BR < 10^{-12}}$} & 
{$ >$ 1 (10)} \\
\hline
 & &  \\
 ${K \rar \mu e}$ & ${ BR < 10^{-12}}$ & $ >$ 3 (4) \\
\hline
 & &  \\
{${K \rar \pi \mu e}$} & {${ BR < 10^{-10}}$} & 
{$ >$ 1 (1)} \\
\hline
 & &  \\
{${ n \lrar{\bar n}}$} &{ $ {\tau > 10^{8}}$ s} & 
{$ >$ 1 (3) { (MSSM)}}\\
\hline \hline
\end{tabular}
\vspace{0.5cm}
\end{center}

Thus we see that TeV scale gravity does not lead to contradiction with
experiment if the condition that virtual BH should have positive mass
and be electrically neutral and non-rotating, is fulfilled. 

TeV scale gravity allows also for succesful, even quite efficinet,
 baryogenesis at relatively low temperatures~\cite{bdf2}. 
All three Sakharov conditions: \\
1) baryon non-conservation,\\
2) deviation from thermal equilibrium,\\
3) large CP-violation in MSM,\\
are much easier fulfilled than in the standard case.

Let us start from CP-violation. It is well known that
CP-breaking in the minimal standard model (MSM) is extemely 
weak. The amplitude of CP-violation is known to be proportional
to the mass differences of all quark families and their mixing
angles, for detials see e.g.~\cite{add-cp}:
\be{ 
\label{cp-suppression}
\epsilon_{CP} \approx
(m_t^2-m_c^2) (m_t^2-m_u^2) (m_c^2-m_u^2)}\nonumber\\ 
{ (m_b^2-m_s^2) (m_b^2-m_d^2) (m_s^2-m_d^2) \,
(J / T^{12})}
\ee
where 
\be{{
J = {\cos\theta_{12}} \cos\theta_{23} \cos^2\theta_{13}
\sin\theta_{12} \sin\theta_{23} \sin\theta_{13}
\sin\delta_{CP} \approx 3 \cdot 10^{-5}.}
\label{jarl}}
\ee
{Thus $ {\epsilon_{CP} \sim 10^{-19}}$ at
$ {T\sim 100}$ GeV.} At lower temperatures the B-nonconserving
sphaleron processes are exponentially suppressed and one cannot 
expect baryon-to-photon ratio larger than $10^{-20}$, while the
observed value is $\sim 5\cdot 10^{-10}$. 

Low scale gravity models lead to nonconservation of baryonic charge
at much lower, than 100 GeV, temperatures because nonconservation of
$B$ takes place simply in decays of heavy quarks and non-perturbative
sphalerons are unnnecessary. According to the estimates of 
ref.~\cite{bdf2}, non-conservation of baryons remains significant 
even at $T\leq 10$ GeV and the amplitude of CP-violation 
(\ref{cp-suppression}) becomes 12 orders of magnitude larger.

One may avoid any suppression of CP violation at high temperatures
if time variation of quark masses is allowed~\cite{nir,bdf2}. In this
case both mixing angles and quark mass differences can be large
in the early universe.

Deviation from thermal equilibrium would be unsuppressed as it
follows from eq. (\ref{H-Gamma}) with $m_{Pl}\sim $ TeV. 

\vspace{0.3cm}
So to conclude:\\
1. Low scale gravity allows for much more efficient baryogenesis
than the standard model.\\
2. In conservative SUSY model with broken \rp 
successful baryogenesis may proceed with (practically) stable proton
and with noticeable neutron-antineutron oscillations.\\
{3. The classical black hole conjecture makes compatible TeV gravity
and low probability of B and L nonconserving processes.} \\
{4. The probability of such rare processes can be quite
close to the existing experimental bounds.}

\end{document}